\documentclass[twocolumn]{article}
\usepackage{graphicx}
\usepackage[total={7.5in,9.5in},top=0.5in,left=0.5in]{geometry}
\usepackage[dvipsnames]{color}
\graphicspath{{.}}

\usepackage[small,compact]{titlesec}

\usepackage[round]{natbib}

\usepackage{amsmath,amssymb}

\title{Feedback Control as a Framework for Understanding Tradeoffs in Biology} 

\author{Noah J.~Cowan$^{1,\ast}$, M. Mert Ankaral{\i}$^1$, Jonathan$^2$ P.~Dyhr, Manu~S.~Madhav$^1$,
  Eatai Roth$^2$,\\Shahin Sefati$^1$, Simon Sponberg$^2$, Sarah~A.~Stamper$^1$, Eric~S.~Fortune$^3$ and Thomas L.~Daniel$^2$}

\date{$^1$Department of Mechanical Engineering, Johns Hopkins
  University, Baltimore, MD 21218, USA, $^2$Department of Biology,
  University of Washington, Seattle, WA 98195, USA, and $^3$Department
  of Biological Sciences, New Jersey Institute of Technology, Newark,
  NJ 07102, USA \\
  $^{\ast}$Author for correspondence (ncowan@jhu.edu)}

\begin{document}
\maketitle

\begin{abstract}
  Control theory arose from a need to control synthetic systems.  From
  regulating steam engines to tuning radios to devices capable of
  autonomous movement, it provided a formal mathematical basis for
  understanding the role of feedback in the stability (or change) of
  dynamical systems.  It provides a framework for understanding any
  system with feedback regulation, including biological ones such as
  regulatory gene networks, cellular metabolic systems, sensorimotor
  dynamics of moving animals, and even ecological or evolutionary
  dynamics of organisms and populations.  Here we focus on four case
  studies of the sensorimotor dynamics of animals, each of which
  involves the application of principles from control theory to probe
  stability and feedback in an organism's response to perturbations.
  We use examples from aquatic (electric fish station keeping and
  jamming avoidance), terrestrial (cockroach wall following) and
  aerial environments (flight control in moths) to highlight how one
  can use control theory to understand how feedback mechanisms
  interact with the physical dynamics of animals to determine their
  stability and response to sensory inputs and perturbations.  Each
  case study is cast as a control problem with sensory input, neural
  processing, and motor dynamics, the output of which feeds back to
  the sensory inputs. Collectively, the interaction of these systems
  in a closed loop determines the behavior of the entire system.  
\end{abstract}


 \section{Introduction}
\label{sec:intro}

The idea that organisms can be understood as a hierarchy of
organizational levels---from molecules to behavior---seems
intuitive. Indeed, a dominant paradigm in biological science involves
a reductionist approach in which a phenomenon at a particular level of
organization is described as a consequence of the mechanisms at a
lower level. For example, the mechanisms underlying behavior might be
described in terms of the activity of a set of neurons, or the
behavior of a single neuron might be understood in terms of its
membrane properties.

However, each level of organization exhibits emergent properties that
are not readily resolved into components \citep{andersonmore1972},
suggesting the need for an integrative approach.  The case for taking
an integrative view is strengthened by considering a fundamental
organizational feature inherent to biological systems: feedback
regulation. Living systems ubiquitously exploit regulatory mechanisms
for maintaining, controlling, and adjusting parameters across all
scales, from single molecules to populations of organisms, from
microseconds to years.  These regulatory networks can form functional
connections within and across multiple levels
\citep{egiazaryantheory2007}. This feedback often radically alters the
dynamic character of the subsystems that comprise a closed-loop
system, rendering unstable systems stable, fragile systems robust, or
slow systems fast.  Consequently, the properties of individual
components (e.g.\ biomolecules, cells, organs), the communication
channels that link them (e.g.\ chemical, electrical, mechanical), and
the signals carried by those channels (e.g.\ phosphorylation, action
potentials, forces), can only be understood in terms of the
performance of the complete feedback control system. Understanding the
role of individual components in the context of a complete feedback
system is the purview of control systems theory
\citep{astrom2008feedback}. Control theory provides a suite of tools
and language for describing biological feedback control systems
\citep{rothcomparative2014}.

An organism comprises a complex patchwork of feedback control systems
that cut across traditional levels of biological organization. Thus,
understanding biological systems requires an understanding of what
feedback can (and cannot) do. Feedback can be used to dramatically
enhance robustness and performance of a system. However its benefits
are not endless: there are inherent, often inescapable tradeoffs in
feedback systems, and control theory provides precise quantitative
language to address such tradeoffs
\citep{middletontrade-offs1991,loozetradeoffs2010,freudenbergright1985}.

Perhaps nowhere are feedback control tradeoffs---such as the intricate
balance between stability and change---more immediately relevant than
in the controlled movement of animals.  While performance measures
such as speed and efficiency are essential to some behaviors, some
measure of stability almost invariably plays a role, as the fastest
animal would have poor locomotor performance if the smallest
irregularity in the surrounding environment was sufficient to cause it
to crash, fall, break, or fail \citep{dickinsonhow2000}. Yet animals
do not seem to adopt the most stable, conservative designs---e.g.\
aggressively regulating convergence rate and bringing the animal to
rest as quickly as possible may be inefficient from a different
perspective \citep{ankaralihaptic2014}.

There is no single definition of ``stability'' or ``change'' that is
ideally suited for all biological systems. In neuromechanical terms,
however, ``change'' can be defined in terms of the responsiveness or
maneuverability of the motor system \citep{sefatimutually2013} to a
sensory stimulus---i.e.\ the ``bandwidth'' of the
system. ``Stability'' on the other hand, refers most generally to the
ability of a system to reject external perturbations, but can also
refer to the ``persistance'' of a system
\citep{ankaralihaptic2014,byl_metastable}.

After providing a short historical perspective on feedback control and
biology, we review a diversity of sensorimotor feedback control
systems. The methods reviewed in this paper are remarkably conserved,
despite categorical differences in species (vertebrates and
invertebrates) and locomotor modalities (terrestrial, aquatic, and
aerial). Indeed, one of the sensorimotor behaviors, the jamming
avoidance response in the weakly electric fish, requires no movement,
but yet enjoys the same basic modeling tools and approaches as used
for the movement based sensorimotor feedback systems.

\subsection{A Brief History of Feedback Control and Living Systems }

\begin{figure*}[htbp]
  \centering
  \includegraphics{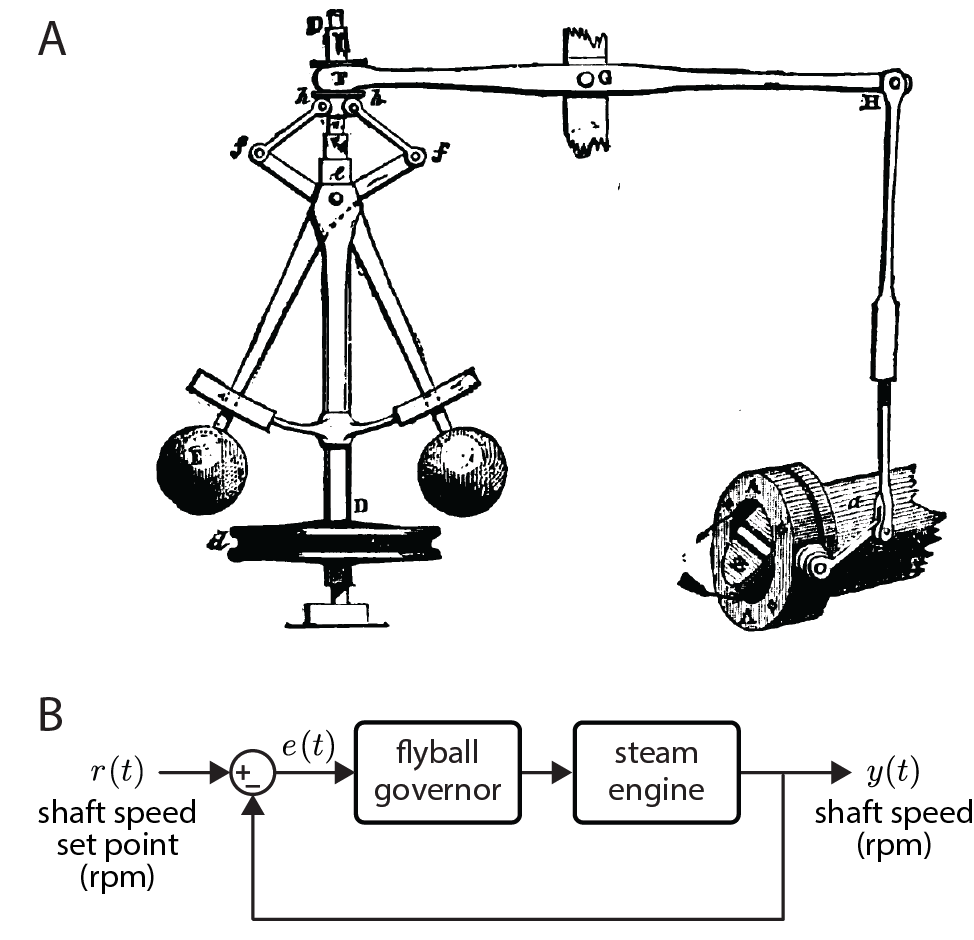}
  \caption{(A) Schematic of the Watt-Boulton centrifugal ``flyable" governor
    (copyright expired; see
    \texttt{http://copyright.cornell.edu/resources/publicdomain.cfm}) and (B)
    a simplified feedback diagram. }
  \label{fig:flyball}
\end{figure*}

In the latter part of the 18th century, James Watt and Matthew Boulton
invented an ingenious device that provided feedback control for the
steam engine. In this device, modeled after similar mechanisms 
used in windmills and millstones \citep{hillspower1996}, the passage of
steam was controlled by a governor (Figure \ref{fig:flyball}) in which
the rotational velocity of the engine's output was ``sensed'' by two
masses suspended from an articulating lever system. As the rotation
rate of the engine increased, the centripetal acceleration raised the
masses closing the throttle to the engine. Thus the output of the
engine was sensed by a physical mechanism that influences the input of
steam into the engine. The governor is an elegant physical
instantiation of a closed-loop feedback system: the spin-rate of the
mass depends on, and controls, the flow of steam through the system.

That concept of a governor---a closed-loop feedback system---has its
tendrils in evolution and ecology.  Indeed, the idea that feedback
plays a central role in evolution owes its origins to Alfred
\citet{wallacetendency1858}:
\begin{quote}
  The action of this principle is exactly like that of the centrifugal
  governor of the steam engine, which checks and corrects any
  irregularities almost before they become evident; and in like manner
  no unbalanced deficiency in the animal kingdom can ever reach any
  conspicuous magnitude, because it would make itself felt at the very
  first step, by rendering existence difficult and extinction almost
  sure soon to follow.
\end{quote}

At about the same time that Darwin and Wallace were forging the
principles of evolution, the concept of the governor in feedback
control extended deeply into dynamical systems theory with James
Maxwell's early contribution ``On Governors''
(\citeyear{maxwellgovernors1867}) in which he lays out the
mathematical formulation for the equations of motion of a governor and
a speed dependent term throttling down the spin rate. The major
contribution of this work was his generalization to a broad class of
closed-loop feedback systems, with the novel idea that the input and
output of such systems are inextricably linked (as they are in living
systems). Maxwell is, in many ways, one of the founding contributors
to the theory of control systems.

Following Maxwell, systems and controls theory takes on a rich history
with a focus on quantifying the stability and performance of dynamical
systems with the seminal contributions of Aleksandr Lyapunov, Harry
Nyquist, and Hendrik Bode (among many others) in the late 19th and
early 20th centuries.  These and subsequent historic contributions of
control theory to engineering domains---from tuning radios to aircraft
controllers, and much more---are summarized in a short history of the
field by \citet{bennettbrief1996}.

The ideas of Wallace and Maxwell can be traced through to living
systems in a host of contributions. For example, at the time of
Maxwell, the French physiologist Claude Bernard developed the idea
(later termed ``homeostasis'') suggested that physiological systems
maintain a constant internal surrounding environment (``la milieu
interieur'') via physiological feedback. Interestingly, glucose
control and its consequences to diabetes was a key example of feedback
developed by Bernard. Indeed, this concept of feedback and control is
a hallmark of the sort of biological systems we study and teach.

In the latter part of the 19th century and early 20th century, the
contributions of dynamical systems theory to living systems began to
form with an initial focus on neural systems and cybernetics.  The
Russian physiologist Pyotr Anokhin for example, developed the idea of ``back
afferentiation'' (feedback) in neural control of sensorimotor systems
and reflexes \citep[for a review, see][]{egiazaryantheory2007}.
Anokhin's ``functional systems theory'' has all the hallmarks of
dynamical systems and control theory we use in cybernetics and in
systems biology today.

While Anokhin was delving into neural systems, reflexes and control,
Norbert Wiener, a pioneer in mathematics and engineering, had begun to
focus much of his attention on the control laws associated with
animals and machines, essentially founding the domain of cybernetics
\citep{wienercybernetics1948}.  Along with the contributions of Ludwig
von Bertalanffy (1969), whose mathematical models of animal growth are
still used today, the fields of cybernetics and von Bertalanffy's
``general systems theory'' gave rise to the burgeoning fields of
systems neuroscience, systems biology, and robotics. Indeed, the sort
of integrative biology we outline in this paper is in every sense
``systems biology'' without the restriction of attention solely to
genetic and molecular scales, but with all the requisite mathematical
underpinnings that began with Watt and Maxwell.

\section{Four Systems that Walk the Proverbial Tightrope}
\label{sec:systems}

We turn our focus to the dynamics and control of motor behavior in
animal systems (neuromechanics) to illustrate basic conceptual issues
surrounding the application of control theoretic analyses that address
the apparent dichotomy between stability and change.  Specifically, we
provide four examples of control theoretic analyses of neuromechanical
systems. For each example, we describe the most important features of
the control system at hand, point out any task-relevant tradeoffs, and
discuss how the organism walks the proverbial tightrope.

A second but equally important goal is to review the application of
control theoretic analyses in interpreting the roles of constituent
components of a biological feedback control system.  The specific
physical details of individual components of the system are most
meaningfully described in the context of the intact control
circuit. With this goal in mind, we review rapid wall-following in
cockroaches, refuge tracking in weakly electric fish, the abdominal
reflex response in hawk moths, and the jamming avoidance response,
again in weakly electric fish.

\subsection{Cockroach Wall Following} \label{sec:systems_cockroach} 
\begin{figure*}
\centering
\includegraphics{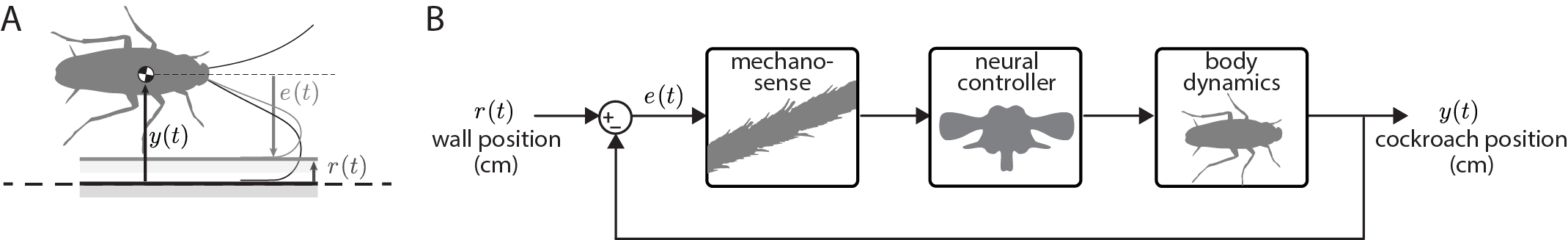}
\caption{(A) Schematic depicting a cockroach following a wall and (B)
  a simplified block diagram representation of cockroach
  wall-following behavior. The reference signal is the position of the wall in
  some global reference, $r(t)$. The difference between the wall and
  the cockroach's position, $y(t)$, is the error signal, $e(t)$. The error is encoded in antennal mechanoreceptors and transformed by the nervous
  system, ultimately causing changes in motor commands that act
  through the animals body dynamics to alter its own position, thereby
  regulating this feedback error to a desired reference point.}
\label{fig:cockroachdiagram}
\end{figure*}

Central to understanding how animals manage stability and change is
evaluating performance in challenging contexts
\citep{dickinsonhow2000}. Crucially, this involves creating
significant deviations from steady-state behavior. Moreover, it
encourages us to consider organisms and behaviors that have evolved in
environments where the pressures, and potential tradeoffs, between
stability and change may be finely balanced.  Locomotion is
particularly challenging when it is difficult to move (surfaces can be
irregular and deformable
\citep{daleyrunning2006,lisensitive2009,sponbergneuromechanical2008,literradynamics2013},
the surrounding fluid can be turbulent and cluttered, or difficult to
sense (some highly visual animals navigate in very low light
\citep{warrantvision2011}), and others behave in complex auditory
\citep{stamperdetection2008} or olfactory landscapes. Behaving in such
environments can be advantageous to avoid competition, predation, and
enable intraspecific interactions.

Despite significant neural resources dedicated to vision, many species
of cockroach navigate low-light and cluttered environments
\citep{bellcockroaches2007}. Rapid locomotion through rough, cluttered
or deformable terrain can render vision unreliable, particularly given
its longer latencies compared to other senses
\citep{franklin2011computational}. Sensor latency can pose fundamental
limits on neural control, and delays are an inevitable part of
processing information in biological wetware
\citep{cowantask-level2006,sponbergneuromechanical2008,elzingainfluence2012,ristrophactive2013,fullerflying2014}.
One strategy cockroaches use is to probe their environment using their
antennae as tactile sensors, mechanically detecting and following
vertical surfaces (or ``walls'') during rapid running (Figure
\ref{fig:cockroachdiagram}A)
\citep{camhihigh-frequency1999,cowantask-level2006,leetemplates2008}.
Thousands of mechanoreceptive hairs lining the cockroach's antennae
are activated on contact and bending
\citep{schaferantennal1973,schallerantennal1978,camhihigh-frequency1999,cowantask-level2006}.
The mechanics of the antenna itself passively maintains the
orientation of the antenna in a ``J'' shape against the wall (Figure
\ref{fig:cockroachdiagram}A) \citep{mongeaulocomotion2013}. The
cockroach runs while maintaining close proximity to the wall and
tracks turns and irregularities in the surface. This behavior enables
remarkably high-bandwidth maneuvers, necessary for maintaining
high-speed performance associated with escape and navigation;
cockroaches reportedly respond to corrugations in a wall with up to 25
turns per second \citep{camhihigh-frequency1999}. Blinding the animals
does not significantly impair performance and they do not require
contact from their body or legs to wall follow
\citep{camhihigh-frequency1999}.

Wall following in cockroaches provides an excellent example of a
smooth pursuit or tracking task
\citep{cowantask-level2006,leetemplates2008}. An external reference
signal, in this case the wall position, is detected by a sensor---the
antenna---and the animal's brain and body cooperate to regulate the
distance from the wall (Figure \ref{fig:cockroachdiagram}). A simple
ethological description of wall-following is that the animal ``tries''
to maintain a certain distance from a surface. This qualitative
description of the behavior is unfulfilling because it is neither
mechanistic---how the animal ``tries'' is not well understood---nor
predictive---we cannot predict how the animal will recover from
perturbations or when its performance will degrade. A classic approach
in neuroethology would be to identify neurons potentially involved in
the behavior and determine what their responses are to a variety of
mechanical disturbances. However, the challenge is that many models
may explain observed patterns of encoding and the relevant mechanisms
may be difficult to identify without rejectable hypotheses derived
from quantification of the animal's dynamics. For example, a
mechanoreceptive neuron may appear to respond to deflections of the
antenna, but its time constants may be too long (or too short) to play
a role the wall-following behavior.  Control theoretic approaches to
understanding the dynamics of both the neural processing and the
body's movement enable testable predictions that inform behavioral,
neurophysiological, and biomechanics experiments.

The control theoretic approach characterizes wall following as a
feedback-regulated behavior (Figure
\ref{fig:cockroachdiagram}B). Using biomechanics models for the body
dynamics---either stride-averaged \citep{cowantask-level2006} or
continuous \citep{leetemplates2008})---we can implement different
hypothesized models for how the nervous system processes the error
signal (i.e.\ the mechanical bending of the antenna) and compare the
resulting dynamics of the whole feedback system to kinematics of
cockroaches following turns in walls. This has led to the discovery
that the animal must encode more than just the position of the
wall. The simplest control model that matches behavior requires the
position and rate of approach of the wall (i.e.\ the lateral velocity
of the wall relative to the cockroach)
\citep{cowantask-level2006}. Such a PD (position and derivative)
controller is ubiquitous in controlled engineering systems. However, while the
latency of the initial turn is low, the turning response persists for
much longer than the stimulus \citep{leetemplates2008}. The system
dynamics filter the sensory input in time. Electrophysiology of the
antennal nerve revealed that the population of mechanoreceptors in the
antennal nerve could encode both position and velocity and that the
neural response even at this first stage of processing is already
temporally matched to the turning response rather than to the stimulus
\citep{leetemplates2008}. In other words, the mechanosensory
processing seems to be tuned to the demands imposed on the control
system by the mechanics and the neural processing delay. A hallmark of
a high-performance control system is the ability to achieve large
responses over a wide frequency range in response to stimuli (change)
without skirting too close to the instabilities that can result from
high-gain, large-latency feedback (stability).

\subsection{Refuge tracking in weakly electric fish} \label{sec:systems_fish}
\begin{figure*}
  \centering
  \includegraphics{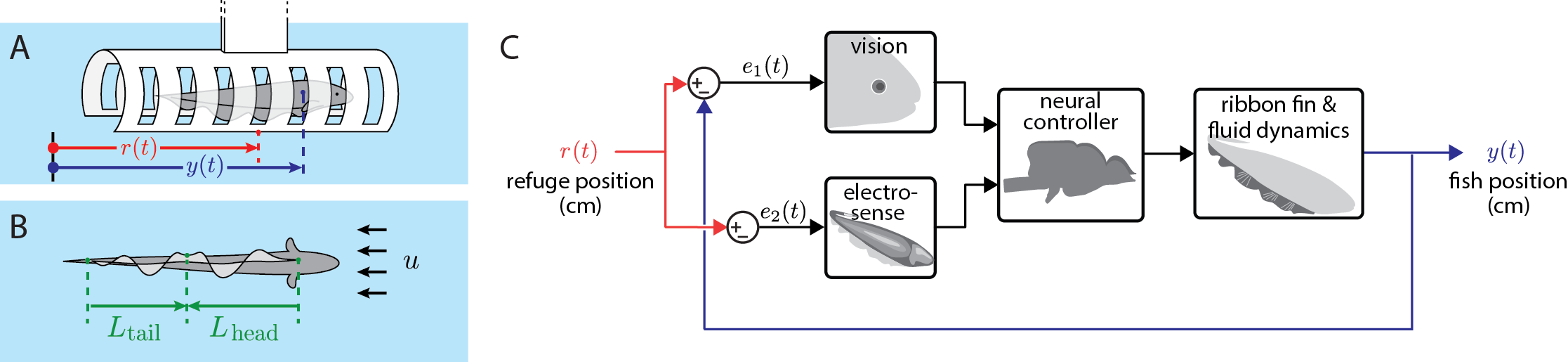} 
  \caption{(A) The knifefish in a moving shuttle. Positions are
    measured from a fixed reference frame to tracking points on the
    refuge and the animal body. (B) A schematic depicting the
    counter-propagating wave kinematics of the knifefish ribbon fin.
    As ambient flow velocity, $u$, increases, fish recruit a larger
    portion of the fin for $L_{head}$, the wave component responsible
    for forward thrust. (C) A block diagram depicting the knifefish
    reference tracking behavior. The moving shuttle provides the
    reference signal, $r(t)$, with the output, $y(t)$, being the
    position of the swimming fish. Parallel visual and electrosensory
    modalities measure the relative position of the shuttle (the
    sensory slip, $e(t)$). The neural controller (CNS) weights and
    filters signals from the sensory blocks and outputs commensurate
    motor commands. Subsequently, these motor commands generate
    movement as dictated the biomechanics of the fish body and the
    body-fluid interaction. }
  \label{fig:fishTracking} 
\end{figure*}

The glass knifefish, \textit{Eigenmannia virescens}, is like an
``aquatic hummingbird'': it hovers in place with extraordinary
precision, making rapid and nuanced adjustments to its position in
response to moving stimuli
\citep{rothstimulus2011,cowancritical2007,roselongitudinal1993,roselongitudinal1993a};
see Figure \ref{fig:fishTracking}A.  Here, we investigate the
integration of locomotor biomechanics \citep[][]{sefatimutually2013}
(Figure \ref{fig:fishTracking}B), multisensory intgration
\citep{stamperactive2012}, and adaptive control
\citep{rothstimulus2011} that enable this animal to balance stability
and change. See Figure \ref{fig:fishTracking}.

As a model organism, weakly electric knifefish are most widely studied
for their namesake, an active electrosensory system. An electric organ
(EO) in the tail emits a weak electric field. Electroreceptors
distributed over the surface of the body (most densely about the head)
detect objects in the near field as small disturbances in transdermal
potential. Using this electrosense in conjunction with vision, fish
perform a wide variety of localization and tracking behaviors. As in
the analyses presented for the wall-following behavior in cockroaches,
we again demonstrate how a control theoretic approach can be used to
quantify and model behavior. For the fish, we further extend the
modeling tool to quantitatively probe the categorical shifts in
behavior and the interplay between visual and electrosensory
modalities (Figure \ref{fig:fishTracking}C).

Weakly electric knifefish hunt nocturnally, their specialized
electrosensory system allowing them to navigate their environment and
localize small prey items in low light
\citep{maciverprey-capture2001}. During the day, they hide from
predators, finding refuge in tree root systems or other natural
shelter. In the laboratory, these fish exhibit a similar
refuge-seeking behavior, hiding in short lengths of pipe, filter
fixtures or any other refuge provided for them. More impressively,
fish smoothly and robustly track their refuge as it is moved (Figure
\ref{fig:fishTracking}A). How do sensorimotor control strategies
differ across this repertoire, in response to different categories of
exogenous motion? And how do visual and electrosensory cues contribute
to these behaviors?

Knifefish are agile. An undulating ribbon-fin runs along the underside
of the body, enabling knifefish to rapidly alternate between forward
and backward swimming without changing body orientation. In
experiments, this remarkable ability is often leveraged to constrain
the behavior to a line of motion, reducing the spatial dimension of
the task to single degree of freedom. Fish were first observed
swimming side-to-side in response to laterally moving plates and rods,
termed the ``following'' response
\citep{heiligenbergelectromotor1973,matsubarahow1978,bastianelectrolocation1987}
and later experiments explored similar behavior in response to
longitudinally (fore--aft) moving refuges
\citep{roselongitudinal1993,rosediscrimination1991,cowancritical2007,rothstimulus2011,stamperactive2012}. In
the case of longitudinal refuge tracking, the fish maintains a
``goal'' position within the refuge, perceiving the error between its
position and that of the refuge and swimming forward or backward to
minimize this positional error (Figure \ref{fig:fishTracking}A and C).

Linear control theoretic tools were used to characterize the
(frequency-dependent) relationship between sensing and swimming in
this smooth-pursuit behavior \citep{cowancritical2007}.  Interestingly,
however, subsequent investigations revealed important deviations from
the linear model proposed in \citep{cowancritical2007}. When comparing
the responses to complex motion (in this case, sums of sinusoidal
trajectories across a broad band of frequencies) and pure sinusoidal
stimuli, fish exhibited performance suggesting qualitatively different
underlying models \citep{rothstimulus2011}. This failure of the
so-called superposition property of linear systems, revealed an
interesting nonlinearity: fish tune their control policy to improve
behavior with respect to the spectral content of the stimulus. While
this adaptive tuning improves the response to a limited class of
stimuli (and these may well be of greater ethological relevance),
performance suffers in response to signals that contain broader
frequency content.

Another tradeoff manifests as a nonlinearity in the context of active
sensing. When visual cues are limited in the environment (e.g.\ in low
illumination or murky waters) fish rely predominantly on
electrosensation.  Under such circumstances, a conflict arises between
the goal of the tracking task---to remain stationary within the
tube---and the needs of the electrosensory system---which requires
active movements to prevent adapting to a constant sensory
stimulus. In low light conditions, the animal performs a rapid back
and forth shimmy superimposed on the tracking response. These
volitional motions are uncorrelated to the refuge motion and can be
discriminated from the tracking response by their frequency
content. While these active oscillations are significant, tracking
performance with respect to the stimulus motion remains nearly
constant \citep{stamperactive2012}. It is posited that these
oscillations serve to enhance electrosensory acuity and permit a high
level of performance in the absence of salient visual stimuli. Fish
employ a strategy in which tracking error and expended energy are
compromised for improved sensing and tracking of the stimulus.

While we present the tracking behavior as a model system for
the study of smooth pursuit and sensory integration, the locomotor
strategy also illustrates a behavioral tradeoff. Knifefish routinely
partition the ribbon-fin into two counter-propagating waves
\citep{sefaticounter2010,ruizkinematics2013}, recruiting the frontal
portion of the fin to generate forward thrust (a head-to-tail
traveling wave) with the rear section (tail-to-head wave) generating
an opposing force. In stationary hovering, these opposing forces
cancel each other. These waves meet at the ``nodal point'' (Figure
\ref{fig:fishTracking}B); simulations on a biomimetic robot reveal
that the net fore-aft thrust is linearly dependent on this kinematic
parameter\citep{sefatimutually2013,curetmechanical2011}. Moreover,
when compared to simpler strategies (e.g.\ recruiting the whole fin and
modulating the speed of a single traveling wave), the use of
counter-propagating waves significantly improves the fore--aft
maneuverability (by decreasing the control effort) and concurrently
enhances the passive stability (stabilization without active feedback
control) by providing a damping-like force to reject perturbations,
thus simplifying the neural control.

\subsection{Moth Abdominal Reflex For Pitch Control}
\label{sec:systems_moth}

\begin{figure*}[t!]
\centering
\includegraphics[width=\textwidth]{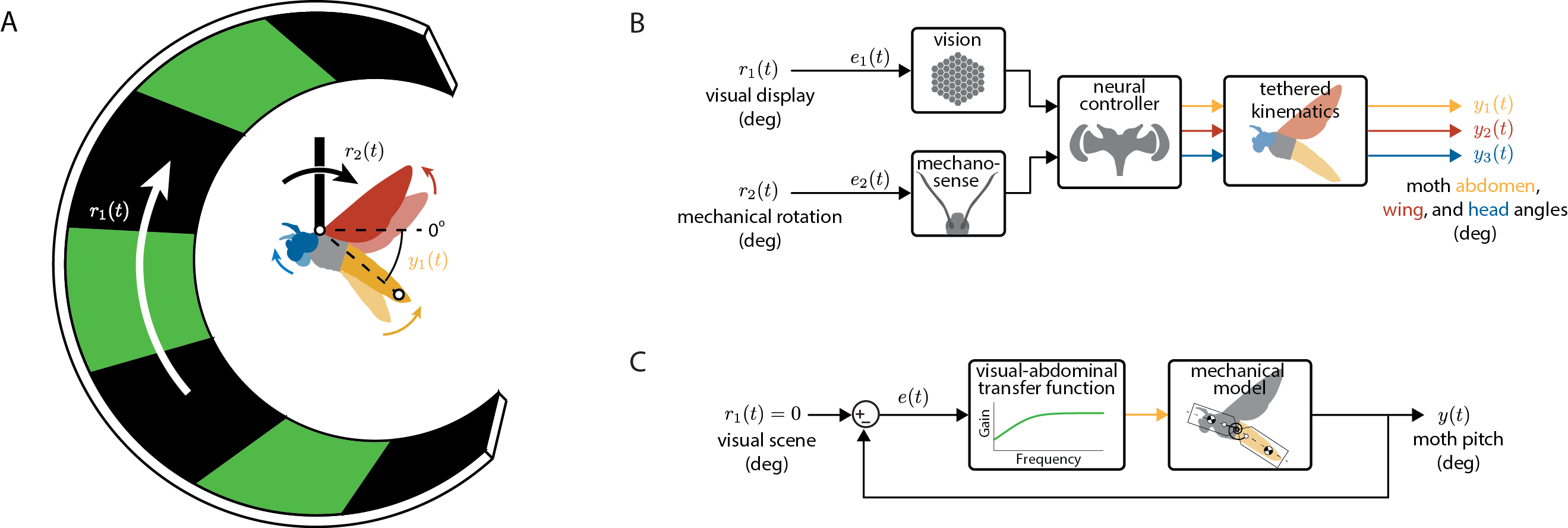}
\caption{(A) Experimental setup for measuring responses to visual
  pitch perturbations in \textit{M.\ sexta}. The moth is attached to a
  rigid tether and placed in a cylindrical LED arena.  During bouts of
  flight the moth is presented with either an isolated visual
  stimulus, $r_1(t)$, by rotating a green and black striped pattern on
  the visual display, an isolated mechanical stimulus, $r_2(t)$ by
  physically rotating the moth and the arena, or a coupled visual and
  mechanical rotation. The moth responds to the rotations by moving
  its head (blue), wings (red) and abdomen (yellow, $y(t)$).  (B)
  Block diagram of the different sensory and motor system known to be
  engaged during open-loop tethered flight.  Error signals, $e_i(t)$,
  represent perceived visual (eyes) and mechanical (antennae) sensory
  information relative to environmental reference signals, $r_i(t)$.
  Sensory systems independently encode the modality specific signals
  and are then fused and processed by the nervous system.  Neural
  commands are relayed to different motor systems to achieve new
  kinematic states.  (C) Block diagram combining open-loop
  experimental data (visual-abdominal transfer function) and dynamics
  models (mechanical model) to estimate the behavior of the
  closed-loop system \citep{dyhrflexible2013}.}
\label{fig:moth_fig1}
\end{figure*}

Here, we examine the problem of active pitch control in the hawk moth,
\emph{Manduca sexta}, as a platform to explore the relationship
between open-loop experiments and closed-loop stability and
maneuverability.  The inherent instability of flapping flight requires
active, feedback-based strategies for control
\citep{wu_floquet_2012,liang_nonlinear_2013}.  This has led to the
evolution of numerous sensory specializations, most notably in the
form of visual and mechanical senses
\citep{ocarroll_insect_1996,pringle_gyroscopic_1948,sane_antennal_2007,taylor_sensory_2007},
that collectively inform the animal about its state in the
environment.  This information, in turn, is used to coordinate motor
systems to direct movement (Figure \ref{fig:moth_fig1}).  

The bulk of research on animal flight control has focused on how the
wings are used to generate and modulate aerodynamic forces.  Much less
attention has been paid to the role of body---or
``airframe''---deformations for flight control.  The hawk moth
displays strong abdominal movements in response to open-loop visual
rotations during tethered flight.  Control theory can provide key
insights about the importance, and possible advantages, of such
movements for controlling flight.

Numerous experimental preparations have been developed for
investigating sensory and motor responses that involve restraining or
confining animals to access physiological signals and to allow for
better experimental control of the sensory inputs available to the
animal. For hawk moths, these include a tethered virtual flight arena
for performing behavioral experiments (Figure \ref{fig:moth_fig1}A) to
immobilized and dissected preparations for electrophysiological
recordings \citep{hinterwirth_antennae_2010,theobald_wide-field_2010}.
However, these types of manipulations dramatically change the dynamic
context of the animal.  The difficulty then is linking physiological
signals and behavioral responses from restrained preparations to free
flight movements, such that causal links can be made between sensing,
movement and flight path changes.

It is here that the analytic techniques of control theory provide
unique affordances for understanding how animals control flight.
Control theory provides a framework for interpreting data from
restrained experimental preparations in the context of the free flight
dynamics via mathematically explicit dynamical models derived from the basic physics and mechanics of flight.  In turn, these studies provide movement predictions
that can be compared to natural flight behaviors.

Hawk moths are accomplished fliers and spend much of their time during
flight hovering in front of flowers while feeding.  Hovering flight is
an equilibrium flight mode that makes the modeling particularly
tractable for control theoretic analyses.  \citet{dyhrflexible2013} took advantage
of these simplified dynamics to test the utility of abdominal
responses for pitch stabilization.  Previous
studies had suggested that abdominal movements were of minor
importance for flight control
\citep{hedrick_daniel_inverse_2006,cheng_mechanics_2011}.  But, absent a control theoretic analysis of free flight, it is difficult to exclude a crucial role for the deformation of the airframe (redistribution of mass).

Free flight is a closed-loop behavior such that movements the animals
make influence subsequent stimuli.  Restraining an animal so that
open-loop responses---in which animal movements no longer influence
the sensory input---can be measured and often generate stronger
responses and more data, simplifying the quantification of the
sensorimotor transform. However, these data can only be interpreted in
their closed-loop context \citep{rothcomparative2014}.
\citet{dyhrflexible2013} combined behavioral experiments, in the form
of open-loop tethered flight responses, with a hovering flight
dynamics model to generate a control theoretic model of closed-loop
visual-abdominal control. See Figure \ref{fig:moth_fig1}.  Using this
model they were able to show that visually evoked abdominal movements
were sufficient for pitch stabilization during hovering flight
independent of any modulation or redirection of the wing forces.

While this work has demonstrated the importance of airframe
deformation for flight control, other actuator systems are clearly
involved. The wings are certainly the most important structures for
flight control, but head movements have an established role in
modulating both visual and mechanosensory information
\citep{dyhrflexible2013, hinterwirth_antennae_2010}.  Understanding
how wing, body, and head movements are coordinated is a promising
future research direction. Future work in this area will require
integrating realistic aerodynamic models with rigid body dynamics to
understand how multi-input control is achieved.  This problem is also
exciting from a multisensory integration standpoint, as hawk moths use
multiple sensory modalities \citep[e.g.\ visual and antennal
mechanosensory,][]{sane_antennal_2007, hinterwirth_antennae_2010} for
flight control.  Furthermore, investigations into the coordinating
multiple motor pathways may highlight the importance of proprioceptive
feedback mechanisms for flight control, an area that has been
relatively unexplored.  The tractability of \textit{Manduca sexta} as
an experimental organism for both behavioral and physiological
studies, coupled with the relatively simple dynamics of hovering
flight, make it a promising model organism for understanding the
sensorimotor processes underlying locomotor control.

\subsection{Jamming Avoidance}

\begin{figure*}[htpb]
\centering
\includegraphics{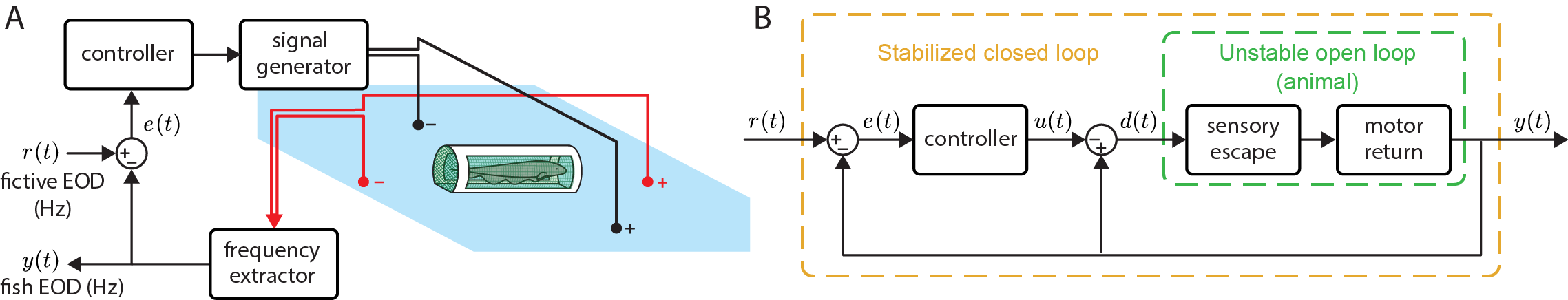}
\caption{(A) Experimental setup to identify the JAR in
  \textit{Eigenmannia}. The fish is placed in a tube in the
  experimental tank, and recording electrodes (red) are used to
  measure its EOD. The EOD is amplified and its frequency is
  extracted. This frequency is fed to the controller which generates
  the appropriate input frequency based on a control law. A signal
  generator outputs a sinusoid at the input frequency, which is then
  played into the tank through the stimulus electrodes (black),
  through a stimulus isolation unit (SIU).  (B) Block diagram
  representation of the same experimental paradigm. The reference,
  $r(t)$, output, $y(t)$, input, $u(t)$, and $dF$, $d(t)$, are all
  frequency signals relative to the baseline frequency of the fish. We
  seek to identify the unstable open loop (green dashed box) using the
  stabilized closed loop (orange dashed box). The $dF$ computation is modeled
  to have a lumped delay. The delayed difference initiates the sensory
  escape, which competes with the motor return to produce the output
  EOD frequency.}
\label{fig:jar}
\end{figure*}

As the previous examples show, control theory is a useful tool for
understanding sensorimotor systems, especially during behaviors which
are robust and repeatable. However, escape responses represent a
behavioral category where the animal actively tries to avoid a
particular sensory condition. From a control systems perspective, the
behavior is transient: it ``escapes'' to the nearest stable
equilibrium. The response is not amenable to perturbation analyses
that we have used so far, and modeling such a response requires a
different approach, such as experimentally ``closing'' the loop.
Here, we re-examine weakly electric fish in the context of just such
an unstable sensorimotor escape behavior.

In addition to sensing the environment for behaviors such as tracking,
described above, the electric organ discharge (EOD) is used for social
communication.  In wave-type fish, each individual produces a
continuous, pseudo-sinusoidal EOD whose frequency and amplitude can
remain remarkably constant for many hours and even days
\citep{bullock1972jar}.  However, if two nearby fish have frequencies
$F_1$ and $F_2$, their electric fields interact to produce a ``beat''
at the difference frequency $|\mathrm{dF}| = |F_1 - F_2|$. When the
two frequencies are within a few Hz of each other, the emergent
low-frequency beat detrimentally interferes with electroreception,
thereby ``jamming'' the ability of the fish to detect obstacles and
prey
\citep{heiligenberg1973electrolocation,bastian1987electrolocation}. Some
species of these fish, particularly those that form social groups
\citep{stamper2010species}, can rapidly change their frequencies to
avoid such interference. This behavior is termed the Jamming Avoidance
Response (JAR).

The neural computation that underlies the JAR in the glass knifefish,
\emph{Eigenmannia virescens}, has been elucidated via a half century
of research \citep{heiligenberg1991neuralnets,fortunedecoding2006}. The
JAR can be predicted on the basis of a single parameter, the $dF$
which can be used to predict the structure of electrosensory
beats. Using parallel receptor systems, the fish encodes amplitude
modulations and relative phase modulations of zero crossings in the
beats to drive a motor response that serves to increase the magnitude
of the $dF$. 

But this is not the whole story. Each individual tends to return to
its own internal EOD frequency set point in the absence of a low
$|\mathrm{dF}|$. The set point can drift over long periods of time,
but over the timescales of the JAR (seconds to minutes) the EOD set
point remains constant. This return to baseline likely serves to
maintain the EOD within the range of frequencies that match the tuning
properties of each individual's own electrosensory receptors. For
example, if a fish were producing a $400$ Hz EOD, its receptors would be
most responsive to EOD frequencies within a range of approximately
$300$ to $500$ Hz \citep{scheich1973electroreceptors}.

How do individuals balance the need for maintaining electrosensory
stability while still being able to rapidly change the EOD frequency
during the JAR?  To address this question,
\citet{madhavclosed-loop2013} modeled the JAR in terms of a low-order
feedback control system that includes both the return to baseline
(stability) and the stimulus-driven escape (change). Parsing this
feedback control diagram (Figure~\ref{fig:jar}) requires understanding
that the JAR operates on frequencies of signals.  Specifically, the
system computes the instantaneous difference between an exogenous
input (conspecific frequency) and the autogenous output (self
frequency) as shown in Figure~\ref{fig:jar}. This difference is
dynamically processed by the CNS which in turn modulates the output,
creating a closed feedback loop.

The challenge in identifying the dynamics of the JAR was twofold. First, the
tendency of the output to diverge from the input renders the system locally
unstable.  Effective system identification techniques rely on analyzing
persistent responses to perturbations; however, in this case, these
perturbations destabilize the system, making it impossible to apply such
techniques. However, this challenge was overcome by stabilizing the behavior
using an experimentally closed loop (Figure \ref{fig:jar}). This stable
closed-loop system was systematically perturbed and the responses were used to
identify a linear model, which describes the unstable open-loop behavior in the
local neighborhood of the baseline frequency of the fish.  Second, predictions
of responses in real-world scenarios requires understanding the global
nonlinear nature of the behavior. A different category of closed-loop
experiments were used to identify a characteristic ``escape curve'', which
serves as the nonlinear signature for the JAR for each individual. Identifying
this curve for each individual allows us to populate all the parameters of the
global model.

In the global model, the computational algorithm of the JAR was
expressed as a competition between the stable motor dynamics (return
to baseline) and the need to adapt to changing social settings
(sensory escape).  Comparatively simple behavioral experiments can now
be used to fit parameters of this model, which can in turn predict
responses to naturalistic or novel artificial stimuli.  For example,
this model captures the asymmetry between rises and falls in EOD
frequency, for which a neural correlate was described previously
\citep{metzner1993twomotor}.  This model could also be used to predict
social interactions between two individuals without considering the
details of each individual's behavioral characteristics. This model
can also form a basis for future work investigating complex social
interactions of three or more individuals, where higher-order
electrosensory envelopes can also drive behavior
\citep{stamperbeyond2012}.

\section{Discussion}
\label{sec:discussion}

We have seen from the above examples that control theory and system
identification tools give us a quantitative framework in which to
interpret comparative organismal studies of locomotion. In cockroach
wall following, we tested hypotheses about neural encoding derived
from the sufficiency of simple control laws and mechanical models. In
the fish swimming and moth flying examples, we understood the
contributions of multiple sensory signals and multiple actuators
(a.k.a. multiple-input, multiple-output or MIMO) to the production of
movement. Along the way we discovered new principles about how animals
are dynamically tuned to their environment. This perspective allows us
to relate the results of closed and open-loop experiments and test
hypothesis about stability and maneuverability. With the jamming
avoidance response, we saw how control theory applies to behaviors
that move not in physical space, but in the frequency space animals
use for infraspecific signaling and communication. These questions of
stability, change, sensing and movement are fundamental to the
emerging field of systems neuromechanics.  We can use the language of
control theory to translate between open- and closed-loop experimental
preparations \citep{rothcomparative2014}, allowing us to relate the
functional mechanisms of individual components to the integrated
performance of the intact, behaving animal, just as the early
pioneers, such as Bernard, Anokhin, and Wiener, envisioned.

\subsection{Closing the Loop from Biology to Control Theory}
Just as control theory affords rich insight into the role of stability
and change in living systems, there is a feedback loop that couples
research on living systems back to the tools we need from control
theory. It is crucial to realize that while control theoretic tools
can enable biologists to tackle challenging problems of great
significance (biomedical, evolutionary, and environmental) the same
can be said of the impact of biology on development of tools in
engineering and control theory. We recall that in fields such as
physics and fluid dynamics, the need for models and mathematical
formalisms continues to spur the development of powerful computational
and analytic methods.

Because biological phenomena are much more complex---chemically,
physically, and organizationally---than inorganic phenomena, a
cause-and-effect understanding of such complex systems will inevitably
foster innovative analytic, computational and technological advances.
Some key examples emerging today include the need for new analytic
methods estimating the dynamics of freely behaving animals
\citep{revzenfinding2011} and reverse engineering biological networks
\citep{kangreverse2013}.
  

\subsection{Integrating Empirical and Physics-Based Models}
\label{sec:mechModel}
The nervous system processes the sensory information for closed-loop
control of task-level locomotion, such as tracking behavior
\citep{cowancritical2007,roselongitudinal1993a}. In control systems
terminology, the mechanical \emph{plant} defines the way motor signals
are transformed into forces and movements, and so discovering the
neural controller
\citep{ekebergcombined1993,fryefly2001,nishikawaneuromechanics2007,rothstimulus2011,miller2012computational}
of a biological system greatly benefits from a task-specific
mechanical model of the underlying locomotor dynamics
\citep{sefatimutually2013,cowancritical2007,cowantask-level2006}.
Low-dimensional, task-specific models for the locomotor mechanics
enable the application of control systems analysis to understand the
neural mechanisms for sensorimotor processing
\citep{blickhan_full.jcp1993,holmesdynamics2006,cowantask-level2006,hedrickwithin2010,tytellspikes2011}. These
simple descriptive mechanical models, sometimes termed ``templates''
\citep{holmesdynamics2006,full_templates}, are essential for
understanding stability and control in biological systems
\citep{blickhan_full.jcp1993,sefatimutually2013,schmittmechanical2000a}.

More elaborate models, sometimes termed ``anchors''
\citep{holmesdynamics2006,full_templates} can facilitate the
exploration of more detailed questions about closed-loop control.
Multidisciplinary approaches integrate computational models and
experiments with biomimetic robots to study the locomotor mechanics in
more details and with higher accuracy
\cite{miller2012computational}. With advances in computing,
high-fidelity simulations have categorically improved our
understanding of various locomotor strategies in different species
\citep{mittalcomputational2004,wangunsteady2004,luoimmersed2008,shirgaonkarhydrodynamics2008,tytellinteractions2010}. On
the other hand, biomimetic robots enable us to experimentally validate
the mechanical models
\citep{wangunsteady2004,lauderfish2007,sefaticounter2012,sefatimutually2013},
and to explore the effect of parameters beyond their biological
ranges, providing insight as to where the biological performance lies
within the range of the wider range of possible mechanical solutions
\citep{curetmechanical2011,sefatimutually2013}.

\subsection{Neurophysiology}

How can the insights concerning the role of feedback in the maintenance of
stability and the control of change at the organismal level be used to decode 
neurophysiological mechanisms used in the brains of animals? 
First we need to determine what we want to learn from and about the nervous
system. In terms of whole animal control, the nervous system is simply one part
of the closed-loop system. In this context, understanding the inputs and
outputs of the nervous system under behaviorally relevant conditions might be
sufficient. The nervous system can remain a black box that is used to better
understand the behavior at the organismal level in its native closed-loop
state.  Alternatively, we may use insights from behavior as a tool for understanding
the properties of the nervous system as a functional unit. Indeed, studying the
nervous system within a closed-loop behavioral task is perhaps the best route
for understanding the functional structure and organization of the nervous
system. In this case, behavior is used to understand the sets
of computations within the black box. 

The central challenge in decoding neurophysiological mechanisms is
that brains are typically composed of millions of independent neurons
each of which may have unique structure and function.
Organism-relevant computations for both maintaining stability and
controlling change are often distributed over thousands to millions of
neurons that act in parallel. Presently we do not understand the
nature of the coding systems that are used in single neurons, and it
is unclear what sorts of dimensional reduction are possible across
populations and networks of neurons.  In other words, there appears to
be no simple or obvious set of \emph{a priori} constraints that
control theory can contribute to decoding the neurophysiological
activity of neurons in the brain. This problem is familiar to
neuroscientists, as one of the long standing challenges in the study
of neural mechanisms is discovering strategies that effectively
translate behavioral observations into feasible neurophysiological
experiments.

This challenge stems in part from the fact that neurons are on the
order of microns to tens of microns in diameter and use tiny
electrical signals. As a result, the vast majority of
neurophysiological experiments have relied on the placement of
microelectrodes into anesthetized and/or immobilized animals or into
neural tissues that have been removed from the animal. Obviously, the
critical organism-level feedback systems that are essential for the
control of behavior are disrupted in these sorts of experimental
preparations. In other words, studies have been conducted in neural
tissues in which the closed-loop control system has been opened by the
elimination of feedback. This is important because it is almost
certainly not possible to extrapolate the neural signals from data
obtained in immobilized animals to make control predictions in the
intact behaving animal (a notable exception is electric fish, in which
certain electrosensory behaviors remain intact in immobilized
individuals \citep{fortunedecoding2006}). These studies in open-loop
preparations cannot capture the dynamics of the closed-loop system,
and further, are likely to be misleading
\citep{szwedencoding2003,cowancritical2007,rothcomparative2014}.

Thankfully, improvements in neurophysiological techniques are now
permitting the widespread recording of neurophysiological activity in
the central nervous systems of awake, behaving animals
\citep{nicolelismethods2008} and with stunningly compact wireless and
battery less technologies coming to play a greater
\citep{thomasbattery2012}. Similarly, recently developed genetic and
optogenetic manipulations
\citep{boydenmillisecond2005,zhangmultimodal2007} can also be used in
animals in which the behavioral control loop remains intact.

\subsection{Characterizing and manipulating internal signals during movement}

One of the strengths of control theoretic approaches is that we can
characterize the relationship between any two signals (neural,
muscular, mechanical, etc.) with the organism as a function of the
underlying neuromechanical system. When we do gain physiological
access we can use the same techniques to relates neural spiking to
movement and sensory feedback to muscle activation. While the examples
in this paper emphasized monitoring motor output while manipulating a
sensory reference single, these signals do not need to be an external
input leading to a kinematic output. Direct alteration of feedback,
either through dynamic manipulation of sensory feedback or by applying
perturbations directly to the constituent neural and mechanical
systems during closed- or open-loop behavior, is the hallmark of the
control systems approach, but is among the least explored experimental
approaches at present \citep{rothcomparative2014}. The ability to
inject both noise and alter neural processing during behavior affords
separation of the contributions of sensors, controllers, and body
dynamics to behavior.

One place where this approach has been used is to identify the role of
individual muscles in the control of movement during posture control,
running, and flight
\citep{sponbergsingle2011,sponbergabdicating2012}. During restrained
or free behaviors these experiments precise altered or "rewrote" the
activation patterns of individual muscles to identify their role in
shaping motor output. In systems where the time constant of the
dynamic response to perturbed motor commands is faster than the
inherent delay transmitting sensory feedback, this also enables the
characterization of open-loop plant dynamics during free behavior
\citep{Sponberg:2011ev}. Alternatively, one can use an open-loop
characterization of the mechanics portion of the systems dynamics. By
replicating the same patterns of input to the muscle but in a isolated
open-loop preparation we can measure the muscles work output (a
``workloop'') \citep{sponbergshifts2011}. Other approaches are
beginning to couple environment forces to \emph{in vitro} muscle
function via artificially closing a feedback loop between a robotic
model and a physiological preparation
\citep{richardsbuilding2011,richardsbio-robotic2012}. From a control
theory perspective the classic \emph{in vitro} experimental approaches
of neuroscience, muscle physiology, and biomechanics are simply ways
to characterize subsystems of the animal (its neurons, muscles, and
body--environment respectively) and each result can be synthesized, in
explicitly quantitative and mechanistic way, back into an
understanding of the dynamics of behavior.

\subsection{Feedback Control in Biological Systems}

Several of the articles in this Special Issue highlight the role of
feedback regulation in the apparent dichotomy between stability and
change across levels of biological organization, from molecular to
ecological.  For example, \citet{grunbaumicb2014} show how ecological
demands can trigger phenotypic changes with complex temporal
dynamics. Variations in the trophic environment triggers a switch
between two phenotypes of radula (``teeth''), ultimately creating a
history dependent pipeline of radulae. From the perspective of control
theory, these temporal dynamics may create a finite-impulse response
filter, allowing the animal to be sensitive to newly available
resources (i.e.  facilitating change) while maintaining a memory of
recently available resources (stability).

At the cellular level, feedback regulation of ATP/ADP is thought to
maintain energy homeostasis, but these homeostatic metabolic systems
may also regulate the development of the respiratory structures
and metabolic pathways that supply oxygen and carbon substrates for
energy metabolism. \citet{greenleepredicting2014} discuss
how the development of insect larval tracheal and metabolic systems
appear to both sustain metabolic performance and plasticity in the
dynamic developmental environment. Such regulation must not, of
course, imperil the longer-term developmental outcomes of organisms
\citep{halestability2014}: if developmental processes are too responsive to the
environment, they could potentially have deleterious effects on adult
structure and performance.

One way to resolve this compromise between stable outcomes and
responsiveness to changing resources may be to incorporate a
combination of feedback and feedforward control. Indeed, this may help
mitigate the tradeoff between preprogrammed developmental
cascades---which may be able to produce consistent outcomes, but are
unresponsive to environmental demands---and tight feedback
regulation---which, while responsive to the environment, may introduce
long-term inefficiencies.  While fault tolerance is a hallmark of
feedback control, even complex feedback control systems are sensitive
to certain categories of failure \citep{csetereverse2002}; such
failures in regulatory networks manifest themselves as disease
\citep{nijhouthomeostatis2014}. In this way, understanding the
mechanisms for feedback regulation in the context of control theory
may be a critical step in the treatment of certain diseases. This
approach will require the development of new quantitative tools, such
as network inference of gene regulatory processes
\citep{ciacciosystems2014}. When such feedback systems are analyzed
using control theory it may enable us to formalize
our understanding of the processes that allow biological systems to
walk the tightrope between stability and change.

\section{Acknowledgements}
This material is based upon work supported by the following grants:
The National Science Foundation under grant nos.\ CISE-0845749
(N.~Cowan), IOS-0817918 (E.~Fortune and N.~Cowan), IOS-0543985
(N.~Cowan and E.~Fortune), BCS-1230493 (N.~Cowan), IOS-1243801
(D.~Padilla), and Postdoctoral Fellowship nos.~1103768 (J.~Dyhr); The
James S.~McDonnell Foundation Complex Systems Scholar Award
(N.~Cowan); The Office of Naval Research under grants N000140910531
(N.~Cowan and E.~Fortune) and N000141110525 (N.~Cowan); The Komen
Endowed Chair (T.~Daniel); The Office of Naval Research grant nos.\
N0014-10-0952 (T.~Daniel); The Air Force Research Lab grant nos.\
FA8651-13-1-004 (T.~Daniel); The Air Force Office of Scientific
Research grant nos.\ FA9550-11-1-0155-MODPO4 (T.~Daniel).

\small
\bibliographystyle{abbrvnat} 
\bibliography{./icballrefs}

\end{document}